\documentclass{article}
\pdfoutput=1

\PassOptionsToPackage{numbers}{natbib}

\usepackage[final]{neurips_2021}

\usepackage[utf8]{inputenc} %
\usepackage[T1]{fontenc}    %
\usepackage{hyperref}       %
\usepackage{url}            %
\usepackage{booktabs}       %
\usepackage{amsfonts}       %
\usepackage{nicefrac}       %
\usepackage{microtype}      %
\usepackage{xcolor}         %

\usepackage{graphicx}
\usepackage{amsmath}
\usepackage{cleveref}
\usepackage{xspace}
\usepackage{textcomp}
\usepackage[export]{adjustbox}
\usepackage{rotating}
\usepackage{wrapfig}
\usepackage{pgfplots}
\pgfplotsset{compat=1.9}

\usepackage{array}
\newcolumntype{P}[1]{>{\centering\arraybackslash}p{#1}}

\usepackage{caption} 
\definecolor{darkred}{rgb}{0.7,0.1,0.1}
\definecolor{darkgreen}{rgb}{0.1,0.6,0.1}

\title{Ownership and Creativity in Generative Models}

\author{
  Omri Avrahami \thanks{Equal contribution.} \\
  School of Computer Science and Engineering \\
  The Hebrew University of Jerusalem \\
  \texttt{omri.avrahami@mail.huji.ac.il} \\
  
  \And
  
  Bar Tamir \footnotemark[1] \\
  School of Computer Science and Engineering \\
  The Hebrew University of Jerusalem \\
  \texttt{bar.tamir@mail.huji.ac.il} \\
}

\begin{document}

\maketitle

\begin{abstract}
Machine learning generated content such as image artworks, textual poems and music become prominent in recent years. These tools attract much attention from the media, artists, researchers, and investors.
Because these tools are data-driven, they are inherently different than the traditional creative tools which arises the question - who may own the content that is generated by these tools?

In this paper we aim to address this question, we start by providing a background to this problem, raising several candidates that may own the content and arguments for each one of them. Then we propose a possible algorithmic solution in the vision-based model's regime. Finally, we discuss the broader implications of this problem.
\end{abstract}
\section{Introduction}
\label{sec:introduction}
Adam Smith argued in his book \cite{smith1976theory} ``The Theory of Moral Sentiments'' that one of the sacred laws of justice was to guard a person's property and possessions. Intellectual property (IP) is a category of property that includes intangible creations of the human intellect. Copyright is a well-known type of intellectual property that gives its owner the exclusive right to make copies of creative work, usually for a limited time.

In statistical machine learning, there are two main approaches: discriminative and generative. Given input $x$ and label $y$, generative classifiers model the joint probability $p(x, y)$ whereas discriminative classifiers model the conditional probability $p(y|x)$. For example, given a dataset of images of cats and dogs, a discriminative model will be able to predict the label (cat or dog) given an image whereas a generative model will be able to generate an image of a dog given a label of a dog. In this paper we will refer to generative models in the broader term - any model that generates new data and not only the mathematically strict definitions (e.g. language models).

Generative models achieved impressive results in recent years in many fields. For example, vision models can generate images with high realism \cite{brock2018large} or even indistinguishable \cite{karras2019style,karras2020analyzing} from real images on specific domains (e.g. human faces). 

There is also a great improvement in the field of Natural-Language Processing (NLP). The  architecture of the transformer \cite{vaswani2017attention}, allowed to use the concept of attention (and specifically self-attention) very efficiently, and generate new and long sequences effectively and also more coherently. BERT \cite{devlin2018bert} applied a Bidirectional Transformer to Language modeling, and presented state-of-the-art results in a variety of NLP tasks, like GLUE (General Language Understanding Evaluation) \cite{wang2018glue} task set, SQuAD (Stanford Question Answering Dataset) \cite{rajpurkar2016squad} v1.1 and v2.0 and SWAG (Situations With Adversarial Generations) \cite{zellers2018swag}. In regards to generating novel text, even hard problems like literature (e.g. poems and stories), GPT-3 \cite{brown2020language} presents very impressive results, in various of languages, and generated even books that were published \cite{book_gpt3}.

Generative models are also used extensively in generating art \cite{park2019semantic, zhu2017unpaired} and generating novel images based on text descriptions \cite{radford2021learning, ramesh2021zero} and novel music (including rudimentary singing) \cite{dhariwal2020jukebox}. Many of these models are being used for novel rendering algorithms \cite{tewari2020state} such as editing talking heads using natural language \cite{fried2019text, yao2020iterative}, generating scenes from unseen novel views \cite{mildenhall2020nerf}, or even performing mathematical theory
the formation, and producing examples, concepts, conjectures, and proofs \cite{colton2008creativity}

Creativity can be defined as the ability to generate novel and valuable ideas  \cite{boden2009computer}. Valuable in a way of interesting, complex and useful ideas. Novelty here has 2 different meanings: psychological and historical. Psychological novelty (or P-Creative idea) is an idea that is novel for the specific creator, whereas Historical novelty (or H-Creative idea) is one that is both P-creative and has never been seen in history before. The classical definitions of creativity are mainly anthropocentric and tend to look for an impact on society \cite{kaufman2009beyond}. There are arguments that we should adopt a less human-centric definition of creativity, in order to get a broader understanding of this concept \cite{kaufman2015animal}.
The absence of a common clear definition of creativity in general, and specifically Computer Creativity (CC) makes it hard to evaluate, and therefore also determine ownership. 

The question of whether computers (and specifically AI models) can be creative in general, and creative as humans in principle, is a big old open question that is part of the bigger question of whether we can relate computers as intelligent entities. Allen Turing designed the Imitation game, known as  ``Turing Test'', in order to determine the ability of a machine to exhibit intelligent behavior \cite{turing2009computing}. Turing suggested a human evaluator that judges a natural language conversation between a human and a machine, that takes place by text only. If the evaluator couldn't determine who is the human participant and who is the machine in the conversation, then the machine passes the test.

A machine learning concept that was adopted in recent years was widely used in state-of-the-art generative models, and reminds this concept of Turing-Test is Generative Adversarial Networks (GANs) \cite{goodfellow2014generative}. This kind of model consists of two main parts: a generator that creates new samples, and a discriminator that tries to evaluate their authenticity. An evolution of GANs are Creative Adversarial Networks (CANs) that try to elevate the creative properties of the new samples from the ones in the training set, and specifically, create a new artistic style \cite{elgammal2017can}. In addition to the aim of making the discriminator identify the generator's sample as a real artwork, it tries to confuse it about the style of the work. This concept skips the hard problem of  evaluating the model's creativity, by using a self-evaluation process, which mimics human evaluation strategy: comparing to other creations seen before, and try to assess whether the new creation is ``on the same level'' so it considers as art, and whether it is distinct enough from them in its style, to be considered as creative, or having a new self-style. The use of this self-evaluation concept encourages the model to  generate creations that are different enough from the training set and to connect more distant semantic spaces to come up with a unique style. It forces the model to generate a creative work.

We start by presenting the problem of ownership that arises from these generative models. We continue with suggesting several candidates for the ownership of such generated content and raise arguments for each one of them. Then we present an algorithmic solution that may help with determining the ownership of generated visual content. We conclude the paper with the broader implications of these models.
\section{Motivation}
\label{sec:motivation}
In this paper, we are going to mainly focus on the models that generate novel imagery and artwork and the models that generate novel texts such as poems and stories. It is important to notice that these models are inherently different than the classic image and text editing algorithms that we being used in the last decades because these models are \textbf{data driven}, which means that they are trained on a large amount of dataset and infers the intrinsic regularities in the data such that they are able not only to trivially generate samples from the training data (memorization) but also, and more importantly, generate \textbf{novel} samples that were not seen in the training data (generalization).

The data-driven nature of these models raises the following question: let us assume that some model was trained on proprietary data, for example, the artwork of a specific artist, and this model is able to generate images from the same image distribution as the distribution of images it was trained on. Who owns the rights for the samples that are drawn from this model? They may be several candidates:
\begin{enumerate}
    \item The owner of the training data.
    \item The person who created/collected the training dataset (if exists).
    \item The developer of the model.
    \item The end-user who generates samples by the model.
    \item The model itself.
    \item No one - public domain.
\end{enumerate}

This question, though may not have a conclusive answer, is influencing many recent generative model based applications: for example, Ganbreeder \cite{ganbreeder} is a collaborative art tool for discovering images, the images are bred by having children, mixing with other images, and being shared. Several artists have that users \cite{cnn_ganbreeder, artmore_ganbreeder} complained about other users ``copying'' the images that were discovered by them using the tool. On the one hand, discovering an image in a model can take time (hence may support the claim the user who found the image first may have a right to the image), on the other hand, the images were solely generated by the model such that any user may find it.

Another very recent tool is Github Copilot \cite{copilot} which is an AI assistant for developers that draws context from comments and code and suggests individual lines and whole functions. The technology behind this model is a language model that was trained on source code from publicly available sources, including code in public repositories on GitHub. There have been claims \cite{copilot_theverge, copilot_recitation} that some of the time this tool emits some code parts that are copied from its training data, these claims are also supported the official \cite{copilot} tool statement ``...We found that about 0.1\text{\%} of the time, the suggestion may contain some snippets that are verbatim from the training set...''.

Yet another complication may arise from tools like GAN cocktail \cite{avrahami2021gan} that are able to train generative models without access to any training data but only to generated samples from other models. It is not clear what is the ownership status of samples that are generated from a model that was trained by another model - the child model ``inherits'' the same ownership status from its parent? Does it get a new ownership status?
\section{Ownership Alternatives}
\label{sec:alternatives}

We've suggested few alternatives of who can owns the Artificial Intelligence (AI) model's creations. 

By the owner of the training data we mean the human author/painter/etc. of the data that the model was trained on. Let us assume that we trained a model which generates poems on Heinrich Heine's poems. A reasonable option is that the new model just mimics the artistic style of Heine, by using his type of metaphors (e.g. by matching words from distant semantic domains, similar to Heine's domains), poem structure, etc. In this case, one can argue that new poems of the model are owned by Heine himself. 

If we take into account that we are in a middle of a new information revolution \cite{rakitov1998information}, it seems plausible that most of the creation, and specifically the artistic creation in future will be produced solely by AI, or with the help of AI machines. In other words, the progress and development of the art world will be depended on AI creation and development.
If every AI creation would be owned by the owner of the training data, there will be no incentive for AI developers, or artists who use AI to create their work, to make new creations. Besides the honor, they wouldn't have an opportunity to make money out of their work (even today, most artists are hardly being paid for their work). In addition, after some years, the creation of artists become a public domain. If most of the artwork would be AI-made, after few decades there won't be any owner to most of the creation in the world, such that the concept of IP will lose its relevance. 

Another alternative for ownership is the user who generated/collected the dataset. Let us assume that while training our poetry model, we used a specific dataset, which consists of only a portion of Heine's poems for example, or poems of several different poets. As we know that a model is highly influenced by the data it is exposed to (this is how it learns), one can argue that the data selection is the main contributor to the model's creative capabilities, hence the dataset creator owns the new poems.
This alternative raises the problem of centralization. Nowadays, a very big amount of digital data is being held by a small amount of big cooperates like Facebook and Google \cite{how_tech_giants_get_data}. Hence, AI models will have to use the data that is owned by these cooperates, in order to be able to get good results. If we apply the proposed alternative, the problem of data centralization would get constantly bigger, as more and more works will be owned by the same small amount of cooperates (a process that feeds itself).

A different optional argument is that the architecture of the model, the choosing of its hyperparameters, and the way it was trained on the data are the crucial properties that make it what it is. In this case, we can consider that the model's developer owns its product. This approach is also aligned with the work-for-hire doctrine \cite{work_for_hire}, where the works of an employee are owned by its employer. In this case, we risk ourselves again with a problem of concentration of control. Once a developer has a trained model and enough computational power, he can easily create numerous of new works. If we take into account the social, cultural, and political influence of art and artists, it could be dangerous (more about broader implications - see section \ref{sec:broader_implications}).

We know that people, and especially artists, are influenced by other works they were exposed to, explicitly and implicitly. There are artistic movements, e.g. Renaissance, Romanticism, Cubism, etc. where we can identify very similar styles among artists of the same movement. Creativity doesn't emerge in a vacuum, and artists are inspired by all the previous works they've seen before. This is a crucial principle in the development of art and artists. Hence, even if a model uses very similar metaphors, shapes or contents, to the ones it was trained on, we can argue that it took its inspiration from them, like a human artist. Another step forward, we can consider the ownership of the model itself on its product. 

What is the meaning of an entity like an AI model to be the owner of its creations? Is it similar to a company that owns the intellectual property on a product has been developed by its employees? Behind the company entity stand humans that actually do the actions, and also people that own the company. These people can use machines that help them to create, but even if their machines create by themselves, the machines don't innovate new things but produce the same product as they were designed to, or programmed to produce. We can argue that the AI model is special in a way that it ``has its own life, or thinking'' once it has been trained. Can we treat the AI model as an intelligent entity, which thinks and act independently, and therefore can own its creations? These are big questions in the field of philosophy and are the kind of questions which Turing Test \cite{turing2009computing} was designed to answer, regarding some machine.
Even if we determine that the model itself should get the IP on its product, we still left we a question of who enjoy the outcomes from it, if exist (e.g. the model's poem is being published on a magazine).

In some cases, like creative works that their copyright term has been expired, the works are considered as a public domain, i.e. no exclusive intellectual property rights are applied on them. As deep learning models are usually trained on a big amount of data, and usually public data, we can consider making their product a public domain. This makes sense because of the publicity of the data it was trained of, and because we can't point on a single distinct owner very clearly, as we saw above.
The problem with this approach is that it may finally make us leaving the concept of ownership, in regards to artificial creation (it isn't a problem specifically for art world, maybe vice versa. The problem is that our nature as persons, and as a result also our society, have become inherently established on ownership nowadays \cite{patane2020me}).

Finally, it seems that we'll have to either change the way we perceive ownership, or make these generative models creating \textbf{with} people, or helping people with their own  creation, so we could only adjust the existing concepts of human ownership, instead of accepting the idea of machines that create and innovate by themselves, and need to get honored for their process.

If models would become more like a tool for creation, we'll be able to grant IP to the user who uses them. As digital artists use different programs that help them to create, these AI models can be kind of helping-tool as well. The main difference is that AI models also create content. Classical tools help a user to create her content, for example by transforming a 2-dimensional drawing of her to a 3-dimensional (3D) one, whereas an AI model can create a 3D drawing from scratch.

A key parameter that can be useful for determining whether the user owns (at least some of the) rights on the creation or not, is the labor-difficulty of creating with the model. If an author use GPT-3 \cite{book_gpt3} for example, only by providing it an initial token, it's hard to believe that she should get honored for that. On the other hand, an artist who uses  Ganbreeder \cite{ganbreeder} for creating images, has a pretty intensive work of searching for getting the final image, even though the image itself is fully generated by the model. In this case, it make sense that the end-user will have rights on the creation that has been generated by the model.
Theoretically, more than one user can find the same image in Ganbreeder, and users can use latent spaces that were found by others. This also encourages the idea of shared ownership, e.g. by a mechanism similar to citations in academic papers. This kind of mechanism can help also with managing the problem that exists in classical art world, the gap between copying and inspiration \cite{inspiration_vs_imitation}. The labor-intensive concept is also aligned with the psychological research about our perception of ownership, and its dependency on creative labor \cite{kanngiesser2010effect}.
The problem with this kind of parameter of labor-difficulty is that it's hard to determine what is hard, what is the hardness threshold and who is authorized to define such parameters?
\section{Possible Algorithmic Solution}
\label{sec:possible_solution}
In this section, we propose a possible solution for the ownership-creativity problem. We are going to focus on the vision-based generative models that generated images. Given a sample that was generated from a model, we want to determine whether the sample is novel (and hence the ownership should be attributed to the model developer/the user of the model - or some combination of both of them) or it is a kind of replication from the training data (and hence the ownership should be attributed to the person who holds the copyright to the data/the specific image that the model replicated).

\subsection{Problem formulation}
Let a generative model $G$ that was trained on a dataset $X$. Given a \emph{specific} sample $s$ that was generated by the model $G$ we want to determine if there exists a reference image $x \in X$ such that $x$ is similar to $s$.

\subsection{Proposed Solution}
In order to find if the sample $s$ exists in some form in the dataset, we should find the k-nearest neighbors using some metric of the distance between the images. Classic per-pixel measures, such as $\ell_2$ Euclidean distance, commonly used for regression problems, or the related Peak Signal-to-Noise Ratio (PSNR), are insufficient for assessing structured outputs such as images, as they assume pixel-wise independence. It was shown \cite{zhang2018unreasonable} that these pixel-wise metrics are not correlated with human judgments of similarity. For example, a blurred version of an image, though considered dis-similar to the original image by human judgment, yields a small distance. Whereas shift by one pixel, that produces an almost identical image as the original (hence considered similar by human judgments) yields a big distance.

In recent years, the computer vision community has discovered that discriminative models, though trained for classification tasks, are useful as an emergent \emph{semantic representation} of images. The de-facto standard model for dealing with images is a convolutional neural networks (CNN) \cite{lecun1995convolutional} and it was shown \cite{zeiler2014visualizing} that CNN operates in hierarchical order such that the neurons in the first layers of the model operate on low-level semantics in the images such as edges and corners and the neurons in the last layers of the model operates in high-level semantics such as objects. It was also shown that the high-level features of these models can be used as a ``perceptual loss'' \cite{johnson2016perceptual} for numerous tasks such as style transfer \cite{johnson2016perceptual} and conditional
image synthesis \cite{dosovitskiy2016generating}.

Later, Zhang et al. \cite{zhang2018unreasonable} have shown that these metrics outperform all previous metrics by a large margin on a dataset of human perceptual similarity judgments (even models that were trained in an unsupervised fashion). They also propose a metric that is based on the perceptual metrics that is called LPIPS score that slightly improves the ``bare'' features by linearly calibrating networks - adding a linear layer on top of an off-the-shelf classification network (VGG \cite{simonyan2014very}).

So, given a sample $s$ we can find the 1-nearest neighbor $x$ of this sample in the training set and calculate LPIPS$(x, s)$ score. If the score is higher than a predefined threshold (that can be arbitrary chosen or calculate using PR/ROC curve of an additional training set of similar an dissimilar images)

\subsection{Limitations of the proposed solution}
Our solution has several limitations: First of all, it is only suitable for images (and can be extended to videos) but it is not clear how to extend it to language and audio model. In addition, the computation time of this solution grows linearly with the dataset, the contemporary datasets such as JFT-300 \cite{sun2017revisiting} contains hundred of millions which may be impractical to calculate. Furthermore, the threshold that is being set of the KNN have a huge influence on the results and there is no ``right'' threshold.
\section{Broader Implications}
\label{sec:broader_implications}
The appearance of  generative models arise broader implications, that go beyond the ownership question itself. There are economical, cultural, social, political implications and more. We'll discuss them all together, as they are inter-connected. 

One of the main fears regarding these models is of people who do this (art)work for their life. Will the AI models replace them, and make them  losing their occupation\cite{towards_will_ai_replace_artists}?
Moreover, artists have a big and important impact on society nowadays\cite{berleant1977artists}: authors' opinions are taken as an ethical measure of ideas and actions, songs and paintings are used in protests against governments, kids are being educated by the same stories and movies, etc.

Do we want our society to be influenced by machines? On the one hand, machines tend to be biased from the bias on the data, which represents, in a way, the bias in our society \cite{qz_microsoft_racist_ai}, so one can argue they are the ``average person'', hence it makes sense to give them a stage. On the other hand, social influencers usually come \textbf{against} social biases, norms, and injustices. We want someone to reflect our problems as a society. In a way models are trained nowadays, it's hard to believe they can catch these positions. In addition, they can be trained to have a specific agenda, so there is the danger of giving one person (the developer, for example), a big power (as theoretically he can create several models with high influence). As we've mentioned before, once you've got a trained model and enough computational power, you can create numerous creations very easily. So the centralization problem become much broader: do we want tech-giants to not only owning most of our personal data, but also having such power to influence our culture, society, politics, etc.? Potentially they'll be able to generate (artificial) artists and content creators in every field and agenda, and design our perception, thinking, choices, etc. They would have an enormous control on people, society and country lives, if we grant the ownership to models' developer exclusively.

Today the government has mechanisms to 'hide', or suppress works that it thinks can danger or harm the country/society (for example by censorship, prevent funding which artists leaned on, bureaucracy, etc.). The creation process is long and tedious, so artists depend on this kind of support. Once we transform the creation process to an automatic one, this dependency is weakened, and the government will have much less power and influence on the creators and their products. 

\section{Conclusions}
\label{sec:conclustions}
We described the impressive development of generative models in the field of AI, and specifically in Computational Creativity: we talked about creativity and demonstrated how it can be produced by AI. 

We presented the question that comes with this development, regarding the ownership of generative models' creations. As these models are data-driven, they create novel samples, and differs from the classical tools that have been used by artist till now. Who have the IP on the novel creations and concepts?

We nominated several alternatives for product's ownership and discussed about positive and negative points we should take into account when we determine whether to accept them or others. We saw that this question refers to the bigger question of the ability of machines to have a self-intelligence.

We showed that a determination on this issue can have broad effects: on the artists themselves (as creators, as people, their motivation, etc.), the art world in general, the developers of such models, our society, culture, politics and more.

It seems from the discussion that a good approach generally would be  distributing the ownership, for several reasons: the creation work is being taking place in much more  cooperative manner while using AI models. In addition, we need both artists and AI developers to have incentive to keep their important work. Finally, it will help managing with the problem of inspiration vs. imitation in art \cite{inspiration_vs_imitation}, and more important, the centralization problem of the enormous power of very few tech-giants, and their ability to influence our life in a variety of aspects.

We proposed the labor-difficulty as a parameter that can help determining the ownership distribution, and an algorithmic solution for ownership-creativity problem in the vision field. The algorithm can help evaluating the novelty of a work, which is a key parameter for creativity and IP. 

We hope artists, AI developers and companies, social organizations and governments will gather all to facilitate a new IP approaches which would be suited to the new developments, and benefit these developments and the people.

\bibliographystyle{plain}
\bibliography{paper_bibliography}

\end{document}